\magnification=\magstep1
\vbadness=10000
\parskip=\baselineskip
\parindent=10pt
\centerline{\bf MATTERS OF GRAVITY}
\medskip
{The newsletter of the Topical Group in Gravitation of the American Physical 
Society}
\medskip
\line{Number 7 \hfill Spring 1996}

\bigskip
\bigskip
\centerline{\bf Table of Contents}
\bigskip
\hbox to 6.5truein{Editorial and Correspondents {\dotfill} 2}
\bigskip
\hbox to 6.5truein{\bf Gravity news:\hfill}
\smallskip
\hbox to 6.5truein{Report from the APS Topical Group in Gravitation, 
Beverly Berger {\dotfill} 3}
\smallskip
\hbox to 6.5truein{We hear that..., Jorge Pullin
{\dotfill} 5}
\bigskip
\hbox to 6.5truein{\bf Research briefs:\hfill}
\smallskip
\hbox to 6.5truein{LIGO Project Status, Stan 
Whitcomb{\dotfill} 6}
\hbox to 6.5truein{General Relativity Survives another Test,
Cliff Will {\dotfill} 8}
\smallskip
\hbox to 6.5truein{Macroscopic deviations from Hawking radiation?,
Lee Smolin
{\dotfill} 10}
\smallskip
\hbox to 6.5truein{Toroidal Event Horizons and Topological Censorship,
Ed Seidel{\dotfill} 12}
\smallskip
\hbox to 6.5truein{Critical behavior in black hole collapse, James Horne
{\dotfill} 14}
\bigskip
\hbox to 6.5truein{\bf Conference Reports:\hfill}
\smallskip
\hbox to 6.5truein{Third Texas Workshop on 3D Numerical Relativity,
Pablo Laguna
{\dotfill} 16}
\hbox to 6.5truein{ICGC-95, Malcolm MacCallum
{\dotfill} 18}
\smallskip
\hbox to 6.5truein{The Josh Goldberg Symposium, Peter Saulson
{\dotfill} 20}
\smallskip
\hbox to 6.5truein{Summer school in Bad Honnef, Hans-Peter Nollert
{\dotfill} 22}
\smallskip
\hbox to 6.5truein{Fifth Annual Midwest Relativity Conference,
Joe Romano
{\dotfill} 25}
\smallskip
\hbox to 6.5truein{Volga-7 '95, Asja Aminova and Dieter Brill
{\dotfill} 27}
\bigskip
\bigskip
\bigskip
\bigskip
\leftline{\bf Editor:}

\medskip
\leftline{Jorge Pullin}
\smallskip
\leftline{Center for Gravitational Physics and Geometry}
\leftline{The Pennsylvania State University}
\leftline{University Park, PA 16802-6300}
\smallskip
\leftline{Fax: (814)863-9608}
\leftline{Phone (814)863-9597}
\leftline{Internet: pullin@phys.psu.edu}

\vfill
\eject
\parskip=5pt
\centerline{\bf Editorial}
 
This is the second issue of MOG as official newsletter of the APS
topical group in gravitation. There have been some changes in the 
way MOG is distributed. We used to have an informal mailing list for
hardcopies of the newsletter. That is being discontinued. From now
on MOG is distributed in hardcopy form to members of the APS topical
group. For those of you on the old mailing list who are not members
we will send a reminder with this issue. Email distribution will 
continue as usual through the gr-qc preprint archive, open to everyone.

In this issue we incorporate an official gossip column ``We hear
that''.  I guess people will have to be careful about what they tell
me at conference banquets. We welcome brief news about people of
the community of interest to others. 

Due to production problems, we could not include an article contributed
by Sam Finn about the Ligo Research Community. It will come out in 
the gr-qc preprint archive.

As usual I wish to thank the correspondents and especially
contributors who made this issue possible. In this issue we say farewell
and thanks to Jim Hartle and welcome  Raymond Laflamme as correspondent on 
quantum cosmology. 

The next newsletter is due September 1st. 
If everything goes well this newsletter should be available in the
gr-qc Los Alamos archives under number gr-qc/9602001. To retrieve it
send email to gr-qc@xxx.lanl.gov (or gr-qc@babbage.sissa.it in Europe)
with Subject: get 9602001 (numbers 2-5 are also available in
gr-qc). All issues are available as postscript or TeX files in the WWW
http://vishnu.nirvana.phys.psu.edu

Or email me. Have fun.

\hfill Jorge Pullin

\bigskip
\centerline{\bf Correspondents}
\medskip
 
\parskip=2pt
\item{1.} John Friedman and Kip Thorne: Relativistic Astrophysics,
\item{2.} Raymond Laflamme: Quantum Cosmology and Related Topics
\item{3.} Gary Horowitz: Interface with Mathematical High Energy Physics and
String Theory
\item{4.} Richard Isaacson: News from NSF
\item{5.} Richard Matzner: Numerical Relativity
\item{6.} Abhay Ashtekar and Ted Newman: Mathematical Relativity
\item{7.} Bernie Schutz: News From Europe
\item{8.} Lee Smolin: Quantum Gravity
\item{9.} Cliff Will: Confrontation of Theory with Experiment
\item{10.} Peter Bender: Space Experiments
\item{11.} Riley Newman: Laboratory Experiments
\item{12.} Peter Michelson: Resonant Mass Gravitational Wave Detectors
\item{13.} Stan Whitcomb: LIGO Project
\parskip=\baselineskip

\vfill
\eject

\centerline{\bf Report from the APS Topical Group in Gravitation}
\medskip
\centerline{Beverly K. Berger, Oakland University}
\centerline{berger@oakland.edu}
\bigskip
\parindent=0pt

$\bullet$ {\bf Elections:}

The first item of news is the successful completion of our first
election. The officers are

\noindent { Chair:} Beverly K. Berger

\noindent { Chair-elect:} Kip S. Thorne

\noindent { Vice Chair:} Abhay V. Ashtekar

\noindent { Secretary/Treasurer:} James A. Isenberg

\noindent { Executive Committee Members-at-large:} James M. Bardeen**, 
L. Samuel Finn*, Leonard E. Parker***,
Frederick J. Raab***, David H. Shoemaker**, Robert M. Wald*

*The number of asterisks denotes the duration of the term in years.

Thanks to everyone who voted!

$\bullet$ {\bf Meeting:}

The official meeting of the TGG will be held during the 1996 APS/AAPT
Joint Meeting, 2-5 May 1996, in Indianapolis, IN. The program of
TGG activities consists of hosting an invited session and jointly 
hosting two additional invited sessions, one with the Division of 
Astrophysics and one with the Topical Group in Precision Measurement and
Fundamental Constants.

{\it * TGG Invited session:}

Clifford Will, "Gravitational Waves and the Death-Dance of
Compact Stellar Binaries''

Frederick Raab, "Progress Toward a Laser Interferometer
Gravitational-Wave Observatory."

Ho Jung Paik , "Spheres -- omni-directional multi-mode
gravitational-wave antennas for next generation."

Matthew Choptuik , "Critical Phenomena in Gravitational
Collapse"

{\it * Invited session with the Division of Astrophysics:}

Peter Meszaros, "Neutron Star Models and Gamma Ray Bursts"

Dong Lai, "Neutron Star Binary Coalescence"

John Friedman, "General Relativistic Instabilities of Neutron
Stars"

Charles Meegan, "Observations of Gamma Ray Bursts"

{\it * Invited session with the Topical Group in Precision Measurement and
Fundamental Constants:}

Francis Everitt, "From Cavendish to the Space Age: Some
Thoughts on the History of Precision Measurement"

Riley Newman, "New Measurements of G"

James Faller, "Precision Measurements with Gravity"

Paul Worden, "Testing the Equivalence Principle in Space"

In addition to these invited sessions, TGG will hold its
first business meeting to be followed by a meeting of the
Ligo Research Community.

Additional information under ``Meetings 
Information'' from http://www.aps.org.

$\bullet${\bf Fellows}

The TGG will be able to nominate at least one of its members
to become a Fellow of the American Physical Society. The
deadline for such nominations to be received by the TGG is
1 April 1996. The procedure for those who wish to nominate
a member of the TGG for Fellowship follows: (1) Insure nominee 
is a member of the Society in good standing. (2) obtain 
signatures of two sponsors who are members of the Society 
in good standing. (3) Submit signed Nomination Form, 
Curriculum Vitae, Biographical Information, Supporting 
Letters prior to the above deadline to: 
Executive Officer, The American Physical Society, 
One Physics Ellipse, College Park, MD 20740-3844, 
ATTN: Fellowship Program 
(see http://www.aps.org/fellowship/fellinfo.html)

If you are reading this newsletter and have not yet 
joined the TGG, you can contact membership@aps.org or follow 
``Membership'' at http://www.aps.org.

Don't forget to browse the TGG home page 
http://vishnu.nirvana.phys.psu.edu/tig

\vfill\eject

\centerline{\bf We hear that...}
\centerline{Jorge Pullin, Penn State}
\centerline{pullin@phys.psu.edu}
\bigskip

$\bullet$ {\it Wai-Mo Suen}
 of Washington University in St. Louis received the 1995
Outstanding Young Researcher Award from the Overseas Chinese
Physicists Association (OCPA).  The award, 
which has been given since 1992, recognizes
achievement by young ethnic Chinese researchers working in North
America and Europe.  The award honored his innovative and
seminal work in the application of computers to solving Einstein's
equations for highly dynamical situations involving black holes and
gravitational waves, particularly his discovery (with Ed Seidel) of a
formulation of an event-horizon boundary condition for numerical
codes.  
He received a \$1,000 check and a certificate during the
``Physics Without Borders'' session of last April's American Physical
Society meeting in Washington. (Thanks to Cliff Will for information).

$\bullet$ {\it Abhay Ashtekar} of Penn State 
was elected an ``Honorary Fellow'' of the Indian Academy of
Sciences.  According to the by-laws: ``This is an international honor
awarded by the Academy for distinguished contributions to
Science. Scientists from all countries are eligible. The total number
of Honarary Fellows can not exceed sixty.''

$\bullet$ {\it Kip Thorne} of Caltech is the winner of the 1996 Julius
 Edgar Lilienfeld Prize of the APS. The citation reads ``For
 contributing significantly to the theoretical understanding of such
 topics as black holes, gravitational radiation and quantum
 nondemolition measurements; for advocating tirelessly the development
 of gravitational radiation detectors; and for conveying lucidly the
 excitement of these topics to professional and lay audiences alike.''
For more information see {\tt http://www.aps.org/praw/96winers.html}
(thanks to Beverly Berger for the information).

\vfill\eject

{\centerline {\bf LIGO Project Status}}
\medskip
{\centerline {Stan Whitcomb, Caltech}}
{\centerline {stan@ligo.caltech.edu}}                                         
\bigskip

Construction continues to move forward rapidly at both LIGO sites
(Hanford Washington and Livingson, Louisiana).  The rough grading
activities at the Louisiana site are underway, and in Washington a
contract to begin the construction of the 8 kilometers of foundations
which will support the beam tube has been awarded.  Our
Architect/Engineering contractor (Ralph M. Parsons Co.) has completed
the preliminary design for the buildings and associated site
development. Parsons is now coninuing with the final design effort, and
the preparation of the bid packages for the construction contracts.

Significant milestones were also achieved on the other major elements
of the LIGO facilities.  A contract for the construction of the LIGO
beam tubes (which connect the vertex and ends of the two arms) was
signed with Chicago Bridge and Iron, the company that performed the
successful demonstration test last year.  They are preparing for full
production of the LIGO beam tubes and plan to begin installation by
fall of this year. The contact for the final design, fabrication and
installation of the remainder of the vacuum system was signed with
Process Systems International, and they have begun intensive design
work.  This design work is scheduled to be complete by summer and will
be followed by fabrication of the hardware to be delivered to the
sites.
 
LIGO helped to organize a second Aspen Winter Physics Conference on
Gravitational Waves from January 15-21, 1996. In addition to the usual
technical interchanges among the experimenters present from the various
groups around the world, there was a special emphasis on data analysis
and the intereraction between experiment and theory in the analysis of
LIGO data.  The conference was also the first meeting of the LIGO
Research Community, an organization of people interested in LIGO
science.  Another meeting of the LIGO Research Community will take
place at the May APS meeting in Indianapolis.

A major effort in LIGO has been to push forward the design of the LIGO
detectors.  After careful consideration, the LIGO Project has made a
working decision to switch its baseline interferometer design to
solid-state lasers operating in the near-infrared in place of Argon ion
lasers operating in the green.  The new lasers are expected to result
in comparable sensitivity and higher reliability in the initial
interferometers.  This decision also defines a clear path for later
improvements to the initial interferometers taking advantage of rapidly
progressing solid-state laser technology, and will permit closer
cooperation with other gravitational wave groups who have generally
adopted solid state near-infrared lasers for their detectors. 

In the R\&D program, investigations of noise on the 40m interferometer
at Caltech  continued. The 40 m interferometer has been converted to an
optically recombined system, and will be converted to a recycled
configuration later this year.  At MIT, the initial phase of research
with a suspended interferometer to investigate optical sources of noise
has been completed. This interferometer, initially configured as a
simple Michelson to emphasize the study of optical sources of noise and
to minimize the amount of time needed to debug other noise sources, has
been fully characterized.  The next step, that of adding a recycling
mirror to increase the optical power incident on the beamsplitter, is
underway.

Further information about LIGO can be obtained from our WWW home page 
at 

\centerline{\tt http://www.ligo.caltech.edu}

\vfill\eject
{\centerline {\bf General Relativity Survives another Test}
\medskip
{\centerline {Clifford Will, Washington University, St. Louis}
{\centerline {cmw@howdy.wustl.edu}
\bigskip

The deflection of light was one of the first great successes of
general relativity, and continues to be an important testing ground.
The original method, whereby the deflection of optical starlight was
detected by measuring the displacement of positions of stars observed
during a total solar eclipse, was replaced during the late 1960's by
the methods of radio astronomy, following Irwin Shapiro's suggestion
[1] that this would ultimately yield higher accuracy.  The idea was to
apply the precision of radio interferometry in differential angular
measurements to monitor changes in the relative angle between pairs of
radio sources, usually quasars, as they pass by the Sun (in radio
astronomy there is no need to wait for solar eclipses).  Although as
much as 10 percent of the total deflection could result from the
refractive effect of the ionized solar corona on the radio waves, this
could generally be accounted for by working at several frequencies
(the coronal effect varies as $f^{-2}$, while the GR effect is
frequency independent), or by not observing the radio sources too
close to the Sun (the coronal electron density falls off rapidly with
distance from the Sun).

Between 1968 and 1975, more than a dozen experiments of this kind were
done, culminating in an accuracy of about 1.5 percent, in agreement
with GR (see [2] for review and references).  The measurements were
taken up again in the 1980s, making use of advances in Very Long
Baseline Interferometry (VLBI), in which the radio telescopes that
comprise the interferometer are separated by transcontinental and
intercontinental baselines.  These advances were made
possible in part by improvements in atomic timekeeping and time
transfer, permitting accurate determinations of the phases of the
radio signals at such widely separated telescopes.  

One effort involved measuring the deflections of 74 radio sources
distributed over the entire sky [3].  The motivation of this program
was geodesy, not GR.  The idea was to use the radio sources as
benchmarks for precise angular measurements to monitor the rotation
rate and axis of the Earth.  Nevertheless, the intrinsic accuracies
(tenths of a milliarcsecond [mas]) were such that the deflection of a radio
source almost anywhere in the sky was measurable.  The grazing
deflection is 1750 mas; for  a source 2.5 degrees from the sun
(the closest source in the sample) the deflection is 184
mas; while for a source at $90^\circ$ from the sun,
the deflection is 4 mas.  Almost 350,000 observations of these sources
were made between 1980 and 1990 using a world-wide network of radio
telescopes.  Two-thirds of the observations were between $50^\circ$ and
$130^\circ$ from the Sun.  What you lose by having small
deflections you gain by statistics, and the reported result for the
overall coefficient  of the deflection, $(1+\gamma)/2$, was $1.0001
\pm 0.001$ [3].  Here $\gamma$ is the PPN parameter whose value is unity
in GR (see [2] for a review of the PPN formalism). 

Recently, results from the more classic method of measuring the
deflection of a single pair of sources very close to the Sun were
reported by Irwin Shapiro and collaborators [4].  Indeed, the sources
were the same quasars, 3C273 and 3C279,  used in many of the original
measurements of the 60s and 70s.  The pair pass by the Sun every
October 8;  3C279 actually is occulted by the Sun.  The  VLBI
observations used two antennas at the Owens Valley Observatory in
California, and two at the Haystack Observatory in Massachusetts, and
were actually made during the fall of 1987.  The quasars were observed
until the corrupting effects of the solar corona became too large,
despite the use of 3 different radio frequencies for most of the
observations, corresponding to a maximum deflection of about 250 mas.
This was the closest to the Sun that such a quasar has been tracked to date.
The result again favored GR, with $(1
+\gamma)/2 =0.9998 \pm 0.0008$.

What are the prospects for improvement?  According to [4], combining
improved observations of specific sources close to the sun with data
from the VLBI surveys of many radio sources could lower the
uncertainty severalfold.  Another option is to use orbiting
observatories with optical interferometers capable of 
microarcsecond precision.
Although there have been a number of
preliminary projects in the U.S. to design and build such observatories, 
NASA seems to have pulled the plug on
them.  A European project called Global Astrometric Interferometer for
Astrophysics (GAIA) has been approved by the European Space Agency as
a possible future mission for the period 2006 -- 2016.  Its stated goal is to
observe 50 million stars with 20 microarcsecond accuracy and to
measure the coefficient $(1+\gamma)/2$ to $10^{-6}$. 

\bigskip
\item{[1]} I. I. Shapiro, Science {\bf 157}, 806 (1967).

\item{[2]} C. M. Will, {\it Theory and Experiment in Gravitational
Physics} (Cambridge University Press, Cambridge 1993), revised ed., p.
167.

\item{[3]} D. S. Robertson, W. E. Carter and W. H. Dillinger, Nature
{\bf 349}, 768 (1991).

\item{[4]} D. E. Lebach, B. E. Corey, I. I. Shapiro, M. I. Ratner, J.
C. Webber, A. E. E. Rogers, J. L. Davis and T. A. Herring, Phys. Rev.
Lett. {\bf 75}, 1439 (1995).

\vfill\eject

{\centerline {\bf Macroscopic deviations from Hawking radiation?}
\medskip
{\centerline {Lee Smolin, Penn State}
{\centerline {smolin@phys.psu.edu}
\bigskip
\parskip=4pt

Most work in quantum gravity assumes that the results of the
semiclassical theory will be reliable until Planck scales are probed.
However, in a recent paper, Jacob Bekenstein and Venceslav Mukhanov
shows that this is not always the case.  This paper, "Spectroscopy of
a quantum black hole"[1] develops earlier proposals of theirs [2,3]
that a black hole be considered a discrete quantum system.
Remarkably, they show that simple assumptions about the spectrum of
black hole states lead to predictions about the spectrum of
evaporating black holes which could immediately be checked, given an
observation of radiation from a black hole of any mass.

The usual derivations of Hawking radiation lead to the prediction that
any black hole will radiate with a perfect thermal spectrum, with a
temperature inverse to the mass, $M$.  (for a neutral non-rotating
black hole.)  A number of authors have speculated that quantum gravity
may lead to corrections to this formula, of the order of
$M_{Planck}/M$ (see, for example [4,5]).  These are, for example,
conjectured to arise if the spacetime geometry is discrete at the
Planck scale.  What is intriguing and surprising about the calculation
of Bekenstein and Mukhanov is that it shows that a particular
assumption about discrete Planck scale structure leads to a deviation
from Hawking's prediction for the spectrum of a black hole that is of
order one, whatever the mass.

The main physical assumption of Bekenstein and Mukhanov[1,2,3] is that
the area of the black hole horizon is quantized, so that
$$
A= n \alpha l_{Planck}^2
$$
where $n$ is an integer and $\alpha$ is a dimensionless constant of
order one.  (Information theoretic reasoning favors $\alpha=4 \ln 2$.)
Bekenstein first proposed that the area of black hole horizons should
come in discrete units more than twenty years ago [2], long before
recent suggestions that something like this (although with a different
spectrum) might be a prediction of a certain kind of approach to
quantum gravity.  Recently, the interest in this kind of assumption
has been increased by Jacobson's remarkable demonstration that the
classical Einstein equations can be derived from statistical
thermodynamics, given a set of assumptions that includes the postulate
that the information enclosed in a boundary is proportional to its
area in Planck units[6].

Given the relationship between area and mass, a direct result is that
the mass spectrum of a spherical black hole is discrete,
$$
M= {1 \over 4}\sqrt{\alpha n\over \pi}M_{Planck}  .
$$
When a black hole evaporates it must make transitions between these energy
eigenstates.  For this reason there is a discrete spectrum of emissions,
with a minimal frequency given by 
$$
\omega_0 = {\alpha \over 16 \pi} {M_{Planck}^2 \over \hbar} {1 \over 2M}
={\alpha \over 16 \pi}  {1 \over 2 G M}
$$
It is facinating to notice that the $\hbar$'s cancel so that the resulting
maximal wavelength is proportional to, and on the order of, the 
Schwarzchild radius.  

If the assumption of integral quantization of horizon area is right then
the consequence is that the emmissions from a spherical black hole
are concentrated in lines with integral multiples of the frequency 
$\omega_0$.  This will differ radically from the Hawking spectrum,
given that the peak frequency of the thermal spectrum is very close 
to the minimal frequency.
($\omega_{peak} \approx 2.82 /8\pi GM $ while with $\alpha=4 \ln 2$
$\omega_{0} \approx \ln 2 /8\pi GM $.)  This means that were 
radiation observed from any spherical black hole, no matter what the
mass, we could immediately distinguish between Hawking's prediction 
from semiclassical methods and the consequences of the hypothesis
of integral quantization of horizon area.

One may question several assumptions about this argument. First of all,
black holes may rotate and absorb and emit charges.  For example, as
photons and gravitons carry away angular momentum, an evaporating
black hole will decay to states with non-zero angular momentum.
One may ask whether this will change the picture of discrete line
radiation.  This is readily checked, and one may conclude that the
result will be a fine structure coming from black holes making
transitions between discrete values of angular momentum as well
as area.  But, as long as the angular momentum is small compared to the
irreducible mass squared (in units $c=G=1$) the resulting 
broadinging of the lines
in wavelength is small compared to $\omega_0^{-1}$ [2].  One
can also investigate further
the statistical properties of the radiation, which Bekenstein
and Mukhanov do[1].  It would also
be interesting to investigate whether different hypotheses about
the quantization of the area, such as those that come from the
loop representation, lead to different predictions about the spectra
of quantum black holes.

Whatever the outcome of this, what is clear is that observations
of radiation emmited by black holes at scales of their Schwarzchild
radii may be sufficient to test hypotheses about the Planck scale.
It has sometimes been said that a black hole serves as a kind of a
microscope which enlarges the high frequency fluctuations near the
horizon to much longer wavelength, which then emerge as the Hawking
radiation.  What Bekenstein and Mukhanov add is that what
is brought into view is, first of all, any discrete structure associated
with the horizon itself.  As a result, the quantum structure of the
geometry of the horizon at the Planck scale is directly observable
in the radiation emmitted at wavelengths of the order of the Schwarzchild
radius of the black hole.

[1]  J. Bekenstein and V. F. Mukhanov, Phys.Lett.B360:7-12,1995, 
e-Print Archive: gr-qc/9505012 

[2]  J. D. Bekenstein, Lett. Nuovo Cimento 11 (1974) 467.

[3]  V. F. Mukhanov,Pis. Eksp. Teor. Fiz. {\bf 44}, 50 (1986) 
[JETP Letters {\bf 44}, 63 (1986)]; in 
{\it Complexity, Entropy and the Physics of
Information\/}, SFI Studies  in the Sciences of Complexity, vol. III, ed. W. H.
Zurek (Addison--Wesley, New York 1990)

[4] T. Jacobson, Phys. Rev. D48 (1993) 728-741, e-Print Archive: hep-th/9303103 

[5]  W. G. Unruh, e-Print Archive: gr-qc/940900.

[6]  T. Jacobson, Phys.Rev.Lett.75:1260-1263,1995, e-Print Archive: 
gr-qc/9504004

\vfill\eject

\centerline{\bf Toroidal Event Horizons and Topological Censorship}
\medskip
\centerline{Edward Seidel, NCSA}
\centerline{eseidel@ncsa.uiuc.edu}
\bigskip

Recent work in numerical relativity has led to the development of
techniques for locating the event horizon of a numerically generated
black hole spacetime[1,2,3].  The event horizon, technically defined as the
boundary of the causal past of future null infinity, is a global
object in time.  However, it turns out that even in a numerical
simulation of finite duration in time, one can quite accurately locate
the event horizon surface, if it exists, in many interesting
simulations.  The basic idea is quite simple: although outgoing light
rays just near the event horizon will diverge away from it going
forward in time, if one integrates backwards in time these same
photons will be attracted to it[2].  Thus if one knows approximately
where the event horizon is at late times in a numerical simulation
(for example by finding an apparent horizon), one can integrate
backwards in time to find a very accurate determination of the
location of the event horizon at earlier times.  What is more, the
horizon surface geometry can be studied, and the horizon generators
themselves can be traced and their properties, including caustic
structures, can be analyzed using this technique.  These ideas of
backward integration and horizon analysis are discussed in detail in [2,3].

Combined with recent developments in numerical evolution of dynamic
black hole spacetimes, including distorted, rotating, and colliding
black holes, these tools open the possibilities for quantitative
studies of event horizons in very interesting spacetimes.  For
example, the event horizons for the collision of two black holes, for
both vacuum and matter spacetimes, and for rotating black holes, has
been studied in a series of papers by different groups[1,2,3,4,5].
The interesting results of [5] are concerned with numerical studies of
rotating black holes, and are a recent example of a developing synergy
between numerical and mathematical relativity that is providing new
insights into general relativity.

In Ref. [5], Shapiro, Teukolsky, and Winicour investigated the
collapse of rotating collisionless matter to form a black hole, using
a numerical code developed at Cornell[6].  A particularly interesting
result, first reported in [1], is that in some cases before the final
Kerr black hole is formed, whose horizon has the topology of a
2-sphere, a topologically toroidal horizon forms, that then evoles into
the expected 2-sphere.  This was the first  example of a
toroidal event horizon.

This result was further analyzed in [5] with the following
consideration in mind: There is a theorem due to Gannon[7] that says
that given asymptotic flatness and the dominant energy condition, an
event horizon of even a non-stationary black hole must be
topologically either a 2-sphere or a torus.  The numerical result of
[3] is clearly consistent with this theorem.  However, Friedman,
Schleich, and Witt[8] showed that any two causal curves extending from
past to future null infinity can be continuously deformed into each
other, creating the idea of "topological censorship".  Following up on
this result, Jacobson and Venkataramani[9] suggested that a toroidal
event horizon might violate topological censorship, since one might be
able to thread the middle of the torus with a causal curve that could not
deformed into a curve that stays "outside" the torus.  Shapiro,
Teukolsky and Winicour investigated how the toroidal horizon
discovered in [1] could be consistent with the topological censorship
theorem of [8].

Aided by a simple flat space model of a toroidal event horizon, they
argue that such a horizon should have a line of "crossovers", where
new generators cross and join the horizon, all the way around the
inside of the torus.  This crossover line is spacelike.  According to
this model the torus, once formed from gravitational collapse,
should close up along the spacelike crossover line faster than light,
as it races to form a 2-sphere.  They then study the actual event
horizon structure of their collapsing, rotating collisionless matter
simulation.  Since at late times this system forms a Kerr black hole
with a spherical topology, by tracing light rays backwards in time
from the final trapped surface, they can find the location of the
surface, and trace its null generators, back through the formation of
the toroidal horizon.  They find that indeed at earlier times,
although the horizon has the topology of a torus, an inner ring of
crossovers is found where photons leave the event horizon (going
backward in time).  They conclude that as this crossover is spacelike,
and closes up to form a 2-sphere faster than the speed of light, no
causal curve can ``link through'' the torus and escape back out to the
exterior region of the spacetime.  Therefore, they conclude, the
numerical result is consistent with all known theorems governing these
systems.

{\bf References:}
\parskip=3pt

\noindent [1] S. A. Hughes, C. R. Keeton, P. Walker, K. Walsh, S. L.
Shapiro, and S. A. Teukolsky,{\it  Phys. Rev. D}, {\bf 49}, 4004, (1994).

\noindent [2] P. Anninos, D. Bernstein, S. Brandt, J. Libson, J.
Mass\'o, E. Seidel, L. Smarr, W.-M. Suen, and P. Walker, {\it  Phys.
Rev. Lett.}, {\bf 74}, 630, (1995).

\noindent [3] J. Libson, J. Mass\'o, E. Seidel, W.-M. Suen, and P. Walker,
{\it  Phys. Rev. D}, to appear, (1996).

\noindent [4] R. A. Matzner, H. E. Seidel, S. L. Shapiro, L. Smarr,
W.-M. Suen, S. A. Teukolsky, and J. Winicour, {\it  Science}, {\bf 270},
941, (1995).

\noindent [5] S. L. Shapiro, S. A. Teukolsky, and J. Winicour, {\it
Phys. Rev. D}, submitted, (1995).

\noindent [6] A. Abrahams, G. Cook, S. L.
Shapiro, and S. A. Teukolsky,{\it Phys. Rev. D}, {\bf 49}, 5153, (1994).

\noindent [7] D. Gannon,{\it Gen. Rel. Grav.}, {\bf 7}, 219, (1976).

\noindent [8] J. L. Friedman, K, Scleich, and D. M. Witt,{\it Phys. Rev.
Lett.}, {\bf 71}, 1486, (1993).

\noindent [9] T. Jacobson and S. Venkataramani, {\it Class. Quant. Grav.},
{\bf 12}, 1055, (1995).

\parskip=5pt

\vfill\eject
\centerline{\bf Critical Behavior in Black Hole Collapse}
\medskip
\centerline{James H.~Horne, DAMTP, University of Cambridge}
\centerline{jhh20@damtp.cam.ac.uk}
\bigskip

Spherically symmetric gravitational collapse of matter is a
surprisingly rich subject. The recent interest in the field was
inspired by the work of Choptuik~[1] in which he studied the collapse
of a massless scalar field. For very weak initial data, the field
bounces away. For very strong initial data, the field collapses to
form a black hole. By using a sophisticated adaptive mesh algorithm,
Choptuik showed numerically that by tuning a one parameter family of
initial data labeled by $p$, he could make an arbitrarily small black
hole (up to computer precision). The precisely critical solution, $p =
p_{\rm crit}$, which is in some sense a zero mass black hole, has a
number of fascinating properties. First, the critical solution seems
to be a universal attractor. All families of initial data approach the
same critical solution at criticality. The critical solution is
discretely self-similar (DSS), so all features are repeated at arbitrarily
small scales. Finally, the mass away from criticality has the form
$M_{\rm bh} = c_i(p - p_{\rm crit})^{\gamma}$, where $\gamma = 0.374$
is a critical exponent, again independent of the initial data.

Not long after~[1], similar results were found numerically in the
study of the axisymmetric collapse of gravitational radiation~[2], and
for the spherically symmetric collapse of a perfect radiation
fluid~[3]. These works did not have the numerical accuracy
of~[1]. They found different scales for the DSS solution, but the same
critical exponent $\gamma \approx 0.36-0.37$. This led people to
conjecture that $\gamma$ was completely universal, independent of the
type of matter.

Recent progress has been made in understanding the scalar field
collapse.  Because of self-similarity, the region near $r=t=0$
contains arbitrarily high curvature in the precisely critical
solution. A continuity argument based on Cosmic Censorship says that
$r=t=0$ should not be visible in finite proper time to an outside
observer.  Unfortunately, Choptuik's initial work was performed in
$r-t$ coordinates, which break down when an event horizon forms.
Hamad\'e and Stewart~[4] looked at the scalar field collapse in
double-null coordinates, which evades this problem. They showed that
the region of high curvature is indeed visible in finite proper time
to an outside observer. Shortly after~[4] appeared, Stephen Hawking
conceded his bet with Kip Thorne about the validity of the Cosmic
Censorship Conjecture. Gundlach~[5] and also~[6] have studied the
critical solution by assuming it is DSS and periodically identifying
coordinates. The solution is still rather complicated.

As was first noticed in the collapse of a perfect radiation fluid~[3],
a great deal of simplification occurs in the critical solution when it
becomes continuously self-similar (CSS) instead of DSS because
Einstein's equations reduce to ordinary differential equations
instead of PDE's. Analytical renormalization group arguments can be
used to study perturbations of the critical solution~[7], and the
eigenvalue of the most unstable mode determines the critical exponent,
with $\gamma = 0.3558$ for the radiation fluid~[7]. This agrees with
the results of numerical calculations, but it was not clear whether it
is significantly different from the scalar field $\gamma$. Similar
analytic calculations indicate that perfect fluids with different
equations of state have different $\gamma$'s~[8], but no numerical
work has been done to show that the CSS solution is actually the
critical solution in those cases.

Another class of solutions which admit CSS solutions are various types
of complex scalar fields coupled to gravity. The CSS solution for the
standard complex scalar field was constructed in~[9], and has a
critical exponent $\gamma = 0.3871$~[10]. However, it was found
analytically in~[10] that the CSS solution is unstable, and they
conjectured that the attractor is actually the DSS solution. This was
confirmed numerically in~[11]. Another type of complex scalar field is
the axion/dilaton field found in string theory. The CSS solution for
the axion/dilaton was constructed in~[12]. We showed numerically
in~[11] that the CSS solution is indeed the critical solution for this
matter, and found that numerical and renormalization group arguments
agree that the critical exponent is $\gamma = 0.2641$. This was the
first conclusive evidence that $\gamma$ is not universal.

In summary, every type of matter studied has either a DSS or a CSS
critical solution. These represent essentially zero mass black holes,
and have singular points at the origin visible in finite proper
time. Numerically, the mass scaling away from the critical solution
has a matter dependent critical exponent $\gamma$, which we now have
the tools to calculate analytically when the critical solution is
CSS. Thus, studying the critical phenomena at the threshold for black
hole formation has given us new insight about the strong field,
nonlinear regime of General Relativity.

I haven't been able to mention all of the work in this field, such as
the lower-dimensional studies or extremal black hole behavior.
Problems for the future include relaxing spherical symmetry, including
quantum effects such as Hawking radiation, and figuring out what type
of matter is physically most important (colliding $D$-branes in
$M$-theory?).

{\bf References}
\parskip=2pt

\noindent [1] M.W.~Choptuik,
{\it Phys.\ Rev. Lett.}\ {\bf 70} (1993) 9.

\noindent [2] A.M.~Abrahams and C.R.~Evans,
{\it Phys.\ Rev.\ Lett.}\ {\bf 70} (1993) 2980;
{\it Phys.\ Rev.}\ {\bf D49} (1994) 3998.

\noindent [3] C.R.~Evans and J.S.~Coleman,
{\it Phys.\ Rev.\ Lett.}\ {\bf72} (1994) 1782, gr-qc/9402041.

\noindent [4] R.S.~Hamad\'e and J.M.~Stewart,
{\it Class.\ Quant.\ Grav.}\ {\bf 13} (1996) 1,
gr-qc/9506044.

\noindent [5] C.~Gundlach,
{\it Phys.\ Rev.\ Lett.}\ {\bf 75} (1995) 3214,
gr-qc/9507054.

\noindent [6] R.H.~Price and J.~Pullin,
``Analytic Approximations to the Spacetime of a Critical Gravitational
Collapse,''
gr-qc/9601009.

\noindent [7] T.~Koike, T.~Hara, and S.~Adachi,
{\it Phys.\ Rev.\ Lett.}\ {\bf 74} (1995) 484,
gr-qc/9503007.

\noindent [8] D.~Maison,
``Non-Universality of Critical Behaviour in Spherically Symmetric
Gravitational Collapse,''
gr-qc/9504008.

\noindent [9] E.W.~Hirschmann and D.M.~Eardley,
{\it Phys.\ Rev.}\ {\bf D51} (1995) 4198,
gr-qc/9412066.

\noindent [10] E.W.~Hirschmann and D.M.~Eardley,
{\it Phys.\ Rev.}\ {\bf D52} (1995) 5850,
gr-qc/9506078.

\noindent [11] R.S.~Hamad\'e, J.H.~Horne, and J.M.~Stewart,
``Continuous Self-Similarity and $S$-Duality,''
gr-qc/9511024.

\noindent [12] D.M.~Eardley, E.W.~Hirschmann, and J.H.~Horne,
{\it Phys.\ Rev.}\ {\bf D52} (1995) 5397,
gr-qc/9505041.

\vfill\eject

\parskip=8pt

\centerline{\bf Third Texas Workshop on 3D Numerical Relativity}
\medskip
\centerline{Pablo Laguna, Penn State University}
\centerline{pablo@astro.psu.edu}
\bigskip

The third Texas Workshop on 3D Numerical Relativity was held in
Austin, Texas from October 30 to November 1, 1995.  The main goal of
the meeting was to report and discuss progress within the Binary Black
Holes (BBH) Grand Challenge Alliance.  The workshop addressed five
areas:

\medskip\noindent$\bullet$ {\it Computational Infrastructure:} 
Choptuik (Texas) reviewed 
the computational strategy of the Alliance.  He pointed out that the
existing and planned numerical codes are remarkably similar with
respect to the basic data structure.  The computational infrastructure
of the Alliance is based on a Berger \& Olinger adaptive mesh
refinement approach (AMR).  Two implementations of AMR with parallel
capabilities are being considered: one is based on Fortran 90/HPF
(Haupt/Syracuse) and the other, called Distributed Adaptive Grid
Hierarchy (DAGH), written in C++ (Parashar \& Browne/Texas).  Choptuik
emphasized that AMR is not a panacea for numerical relativity
problems; proper unigrid construction is the key to success.  Other
talks addressed programming and problem solving environments, Web
tools, I/O and visualization.

\medskip\noindent$\bullet$ {\it Hyperbolic Formulations:} 
Within the last year, it became apparent that Hyperbolic formulations
of Einstein's field equations have the potential of providing a
natural arena to implement apparent horizon boundary conditions and
facilitate the extraction of radiation.  An overview talk of
hyperbolic reductions for Einstein's equations was given by Friedrich
(Postdam) [1,2].  In general terms, depending on the starting point,
there are two types of reduction formalisms: those based on the ADM
equations and those whose starting point are the Bianchi identities.
The BBH Alliance is currently considering two hyperbolic approaches in
the development of numerical codes (Bona et al. [3] and Choquet-Bruhat
\& York [4]).  Reviews of these approaches were presented by York
(North Carolina) and Masso (NCSA).  Both speakers stressed that a
suitable hyperbolic formulation should allow for an arbitrary choice
of shift vectors; furthermore, from the numerical point of view, it is
convenient to be able to write the system as a flux conservative,
first order, symmetric hyperbolic system.

\medskip\noindent$\bullet$ {\it 3D Simulations:} There are three groups 
within the Alliance developing 3D black hole evolution codes: NCSA,
Texas and Cornell. Seidel's (NCSA) presentation reported the progress
in simulating distorted black holes and, in particular, the first 3D
head-on collision of black holes.  Suen (NCSA/Wash. U) addressed a
critical aspect in the evolution of black hole spacetimes, namely
inner boundary conditions when black hole singularities are excised
from the computational domain.  The Texas effort on evolution codes is
based on the standard ADM equations. Correll (Texas) presented results
from evolving Schwarzschild data in Novikov coordinates.  Finally, the
Cornell group presented the status of its 3D code based on
Choquet-Bruhat \& York hyperbolic formulation.  The code is still
under construction; however, preliminary results from evolving
Schwarzschild data seem significantly promising.  Baumgarte (Cornell)
presented an interesting and useful approach to finite differencing
complicated tensor equations, such as those present on the
Choquet-Bruhat \& York hyperbolic approach.  This method takes
advantage of Fortran 90 pointer aliases to produce ``clean" codes.

\medskip\noindent$\bullet$ {\it Outer Boundary:} A series of talks addressed
outer boundary conditions for the evolution codes and approaches to
radiation extraction.  The outer boundary infrastructure consists of
three components: (1) identification of a world-tube along which data
is extracted from the interior evolution, (2) exterior evolution and
computation of the asymptotic waveforms, and (3) interpolation of data
from the exterior evolution to the outer boundary of the interior
evolution.  For the exterior evolution, two approaches are currently
under consideration: A perturbative method, which was reviewed by
Abrahams (North Carolina) and a characteristic approach under the
direction of Winicour (Pittsburgh).  Bishop (South Africa) presented
estimates of the computational efficiency of extraction and matching
using the characteristic approach.

\medskip\noindent$\bullet$ {\it PPN and Perturbation Theory:} Finn (Northwestern) considered
issues regarding the matching of binary black hole initial data with
PPN calculations of the inspiral phase.  An important aspect in the
numerical simulation of black hole collisions is the physical
interpretation of the initial data, in particular, its radiation
content.  At the final stages of coalescence, during the black hole
ringdown, the work by Price and Pullin [5] has proven to work
remarkably well [6]. Pullin (Penn State) presented current attempts to
generalize this work to cases in which the perturbations are about
initial data containing close black hole binaries.

\bigskip
\leftline{\bf References:}

\item{[1]} H. Friedrich, {\it Proc. Roy. Soc. A}, {\bf 375}, 169, (1981). 

\item{[2]} H. Friedrich, {\it Commun. Math. Phys.}, {\bf 100}, 525 (1985).

\item{[3]} C. Bona, J. Masso, E. Seidel, J. Stela, 
{\it Phys. Rev. Lett.} {\bf 75},
           600, (1995).

\item{[4]} Y. Choquet-Bruhat, J.W. York, {\it C.R. Acad. Sci. Paris} (1995).

\item{[5]} R. Price, J. Pullin, {\it Phys. Rev. Lett.} {\bf 72}, 3297 (1994).

\item{[6]} A. Abrahams, G. Cook, {\it Phys. Rev. D} {\bf 50}, 2364, (1994).

\vfill\eject
\centerline{\bf ICGC-95, Pune, India, Dec. 13-19} 
\medskip
\centerline{M.A.H MacCallum, Queen Mary \& Westfield College, London}
\centerline{mm@maths.qmw.ac.uk}
\bigskip

\parskip=8pt
This was the third 4-yearly meeting in the ICGC series and was held at
the Inter-University Center for Astronomy and Astrophysics, which
proved to be a magnificent place for such a gathering. In this short
account I can only mention a few of the many interesting points from
the talks and workshops.

The Organizing Committee, inspired by its Chairman, T. Padmanabhan,
had chosen not to make this an all-purpose conference, but to
concentrate on four themes: Cosmology, especially observations;
Quantum Gravity; Gravitational Radiation; and Classical General Relativity.

The real universe, i.e.\ what we observe, was discussed in plenary
lectures by Richard Ellis, Malcolm Longair and Jim Peebles. Ellis,
speaking on lensing, showed that the Hubble Space Telescope gives us
such good data on lensed images that we can in some cases tightly
constrain the lens.  It also allows us to investigate high-redshift
galaxies without selection bias: there may already be evidence of
flattening of the source counts. Current estimates of $\Omega$ from
this data are in the range 0.4 to 0.6. He also said that recent
observations of supernova of type 1A give $q_0 = 0.3 \pm 0.3$ give a
good reason for thinking $\Lambda = 0$. Longair displayed HST data to
show that the most active epoch of galaxy formation was at $z = 2-3$,
around the maximum of the galaxy number evolution: there were some
striking images of star-formation apparently triggered by radio jets.
Peebles argued that various evidence suggests that galaxies trace
mass, and that $\Omega \approx 0.2$. In a workshop, Rachel Somerville
explained why the often-quoted figure of Davis and Peebles for galaxy
velocity dispersion was an underestimate.

Turning to the imaginary universe, i.e.\ the theorists' models,  George
Ellis gave a number of recent important applications of classical GR
in cosmology, such as the almost-EGS theorem,  on the relation of
almost-isotropy of the microwave background to the almost-isotropy of
the universe, and the effect of focussing on inferred distances at the
last scattering surface. Padmanabhan gave a nice review of
nonlinear gravitational effects in structure formation, and Katz
discussed the integral constraints, giving a nice derivation of the
Traschen form from Lagrangian superpotentials.

In the workshops, Bagla noted that the orthodoxy of the 80s (inflation
with cold dark matter, $\Omega=1$ and $H_o \approx 50$) was now in
trouble, which I regard as a very healthy development, it having for
too long been the case that people (like me) who did not accept this
view uncritically were simply ignored. Padmanabhan criticised another
tendency in the field, namely that workers tend to ignore the results
of large-scale structure calculations if they conflict with other
arguments, rather than  worrying about a genuine discrepancy: perhaps all
those with an interest in cosmology should try making such
calculations so they get an idea of the reliability.  I also noted, in
Sasaki's talk, the advent of inflation theories with low $\Omega$.
Longair and Peebles declared themselves ``hard-line big-bangers'',
meaning they preferred to focus on the observational evidence rather
on theoretically preferred values or models, and Longair in particular
stressed that some of the parameters often quoted are based on very
little data and that one should not rely on true values being
inside $1\sigma$ error bars.

In quantum gravity, Halliwell gave an excellent talk on the
interpretation of quantum theory based on the decoherence concept,
Varadarajandescribed canonical quantization of dilatonic black holes,
in work aimed at further understanding of the Hawking effect.
Torre showed the internal time concept still has possibilities, and
Pullin discussed some issues related to the new variables. I did not
attend the quantum gravity workshops, but they will be reported in
the forthcoming proceedings: they included papers on non-standard
approaches, on semiclassical cosmology, and on quantum mechanics in
non-inertial frames.

Shoemaker and Vinet gave us some insight into the subtleties of
building the LIGO and VIRGO detectors, and the extreme care needed to
reach high sensitivites and avoid noise sources. The planning and
engineering involved are awe-inspiring: the facilities are intended
for more than 20 years use. One-third of the total cost is in the
vacuum tubes (including the concrete casing necessary to protect the
tubes from the U.S. public's gun mania)! All scales, financial,
spatial, temporal, accuracy, noise levels, were impressively large or
small as appropriate.  In the workshop Blair emphasized that bars will
continue to be important: their sensitivities around 1 KHz will
improve to remain comparable with first-generation LIGO, while an
array of, say, 30 small bars could be a very effective instrument in
the 4-7 kHz range looking for collapses of stars below 5 solar masses.

Sources were also considered, in plenary talks by Blanchet and Finn,
which related to the very detailed calculations of wave generation,
some of which were discussed in more detail in the workshop. There
also, Blair noted the potential importance of a stochastic background
from supernovae.

Classical relativity featured in Friedman's talk on topological
censorship, though part of the argument rested on quantum theory, and
Seidel gave a good talk on the present and future of numerical
relativity. There was also an additional day devoted to celebration of
40 years since the publication of Raychaudhuri's paper containing his
famous equation, and attended by Prof.\ Raychaudhuri himself, who gave
some interesting background on its publication history and the
reactions of referees. The speakers, on various aspects of
singularities, collapse and black holes, were Brill, Clarke, Joshi and
Szekeres. My own workshop talk set out some unsolved problems, while
Senovilla reminded us that not all cosmologies were singular and
Tavakol emphasized the importance of robust predictions if models were
to be meaningful.

Seidel noted that there was scope for work on numerical relativity by
small groups in association with the main centres.  To me this
emphasized the benefits to our community from the improved
communications of which this newsletter is part, but those do not
detract from the value of personal meetings, for which ICGC-95 proved
an excllent opportunity: the organizers are to be congratulated.

\vfill\eject
\centerline{\bf The Josh Goldberg Symposium: Five Decades of 
Relativity at Syracuse}
\medskip
\centerline{Peter Saulson, Syracuse University}
\centerline{saulson@suhep.phy.syr.edu}
\bigskip

	A celebration marking Josh Goldberg's seventieth birthday was held
on December 2, 1995 in the Physics Department at Syracuse University. Six
speakers discussed various aspects of general relativity, quantum mechanics,
and their intersection before a standing room only crowd of present, recent,
not-so-recent, and honorary members of the SU Relativity Group. Attendees
had assembled from as far away as Britain, Poland, and Indonesia.

	Peter Bergmann, the founder of the Relativity Group at Syracuse
and Josh Goldberg's original mentor, led off the morning session with a 
talk called "The Quest for Quantum Gravity". He reviewed the basic 
principles of the canonical quantization program. He explained that
this program has occupied so much of the attention of the Syracuse
group because it does not require the introduction of a background
metric, and thus respects the spirit of general relativity. No idealogue,
however, he concluded by emphasizing that all viable approaches to
the problem of quantum gravity should be pursued, a view that he had
held for over three decades.

	Jim Anderson devoted his talk to the question "What is an
Equation of Motion?", a subject to which Josh contributed substantially.
Anderson's argument was that the Einstein-Infeld-Hoffman formalism is the
only good approach in the context of general relativity, in spite of the 
fact that today the subject is considered obscure by many. Rather than 
providing a universal general-relativistic equation of motion, EIH gives a 
prescription for deriving approximate equations of motion tailored to 
different problems. Especially interesting to this reporter, an 
experimentalist, was Anderson's use of the method of multiple time scales 
to propose that time measures defined by 1) the dynamics of gravitationally 
bound systems, 2) electromagnetically-based clocks, and 3) the expansion 
of the Universe would each disagree with the other two, if they could be 
compared with sufficient precision.

	The morning session concluded with a talk by David Robinson on
the subject of "Lagrangians, Hamiltonians, and Einstein's
Equations". His thoughts were motivated by a feeling of disappointment
that, after 80 years, we have as yet no algorithmic way of solving
Einstein's Equations, even for the vacuum case. Robinson believes that
what is needed is a {\it geometric} reformulation of Einstein's
equations. In the half flat case, the existence of a hyper-K\"ahler
structure does this job. He expressed the view that the use of spin
3/2 fields might lead to a similar description of the full vacuum
equations. He concluded by indicating several recent developments that
suggest that a solution to the problem may in fact be close at hand.

	Ted Newman's contribution, leading off the afternoon, was called
"Light Cones and Quantum Gravity". In it he presented his work on 
re-expressing general relativity in terms of the geometry of null surfaces.
This unorthodox point of view can also give a new point of view on
linearized gravity. Newman's hopes for this approach come from the fact
that it leads to a "fuzzy" description of space-time, even at the classical
level. Perhaps this feature could be useful in the development of a full
quantum theory of gravity.

	Bob Geroch treated the attendees to "Three Tales of the Initial
Value Formulation". He noted that the initial value problem is one of the
central paradigms of physics, but that we nevertheless have an incomplete
understanding of the circumstances under which it can be well posed. 
He illustrated this point with three disparate stories, described as a
pair of mysteries plus one sitcom. He pointed out the difficulties
in describing dust, an elastic solid, and gravity in a way that
can satisfy the requirement that the equations of motion be symmetric 
and hyperbolic. 

	The final formal talk of the day was given by Abhay Ashtekar. 
Reporting on work that made up the thesis of his student Troy Schilling,
Ashtekar described a novel view of "The Geometry of Quantum Mechanics."
By comparing the mathematical structures of classical mechanics and quantum
mechanics using the insights of the study of geometry, Ashtekar showed how
one can recast quantum mechanics as a special case of classical mechanics.
This geometric reformulation, although equivalent to the usual algebraic
one, opens doors to non-linear generalizations of quantum mechanics. As
Penrose has suggested such generalizations may play an essential role
in quantum ravity. 

	After a pleasant dinner, the day was concluded by remarks from
John Stachel. He placed Josh's career in the context of an apostolic
succession, in which Peter Bergmann played the role of the rock upon whom
Einstein founded his church of general relativity. Stachel underlined
the great value of the clarity of reasoning and writing in Josh's early 
work, citing it as the single good way for those outside the inner circle of
relativity to come to understand the progress that was being made. Many
of the participants then rose to give testimony on the important role
that Josh had played in their lives and careers. The laudatory mood
was interrupted only by Ted Newman, who told a tale linking Josh
to an international incident involving the attack on Pearl Harbor and the 
sinking of the Titanic, that can, we trust, be consigned to the apocrypha.
\vfill\eject

\centerline{\bf Relativity and scientific computing}

\smallskip
\centerline{\bf A summer school in Bad Honnef, Germany, Sept. 18 -- 22,
               1995} 
\medskip
\centerline{Hans-Peter Nollert, Penn State}
\centerline{nollert@phys.psu.edu}

\bigskip
General Relativity is still regarded by many as something done with
pencil and paper, leading to mathematically exact results which may,
or may not, have physical meaning. However, computers have now become
an important tool in general relativity as well. This summer school
was devoted to giving young scientists and advanced students
specializing in relativity a comprehensive overview over the use of
computers in General Relativity.

The summer school was organized by the `Gravitation and Relativity
Theory' section of the German Physical Society (DPG), together with
the German Astronomical Society (Astronomische Gesellschaft). It's
main financial support came from the WE-Heraeus Foundation, with a
contribution from the Graduate College `Scientific Computing'
K\"oln/St.\ Augustin.  It took place in the physics center of the
Deutsche Physikalische Gesellschaft, an impressive historical building
surrounded by a small park, located in Bad Honnef, a town on the Rhine
river just south of Bonn.

There are three main areas of computer application in GR: Numerical
techniques for solving the field equations (and possibly matter equations),
computer algebra, and visualization for results and diagnostics. 
Lectures on numerical applications included a review of the commonly
used ADM (3+1) formalism, alternatives to this formalism, and standard
numerical techniques used in this context. 
Computer algebra was covered by an overview over existing software, by
in-depth lectures on specific packages, and by applications, such as solving
partial differential equations. 
A general review of scientific visualization was provided, and specific
aspects arising in connection with special and general relativity were
discussed. In addition, relativistic visualization could be seen ``in
action'' as a part of many other lectures as well.

{\bf Ed Seidel} kicked off the lectures with an exciting review of the ADM
(3+1) formalism and the conceptual as well as technical questions
involved, such as slicing and gauge conditions, boundary conditions,
initial data problem, extraction of graviational waves, or locating
horizons. In his second lecture, he concentrated on results obtained
by the Grand Challenge alliance, and by NCSA in particular, ranging
from the evolution of a single black hole to the collision of two
black holes in 3D to spacetimes including Brill waves.

{\bf David Hartley} compared the characteristics of many available
computer algebra tools. Details, such as whether expressions are
evaluated immediately and/or recursively after the substitution of a
variable, can make a considerable difference in a specfic case. A
variety of examples provided some indication which system can be used
for which type of problem.  Hartley also delivered a lecture authored
by {\bf Eberhard Schr\"ufer}, who could not present it himself due to
illness, on the differential geometry system EXCALC and its
applications on relativistic physics and fiber bundles.

Making the invisible visible - this definition of visualization
provided the guiding principle for the lectures by {\bf David
Kerlick}. He discussed the mathematical and technical basis of
computer graphics, summarized available visualization techniques and
environments, and demonstrated scientific applications. An evening
show of visualization videos spanning a wide range of fields was very
much appreciated by the audience.

{\bf Harald Soleng} introduced the Mathematica packages CARTAN and
MathTensor for tensor analysis. CARTAN is designed for tensor
component calculations in Riemann-Cartan geometries, while MathTensor
provides a framework for indicial tensor manipulations. 

{\bf Carles Bona} discussed the definition and the conditions for
hyperbolicity of Einstein's evolution equations, and the relevance of
flux-conservative systems in Numerical Relativity. He reviewed
finite-difference numerrical methods and introduced
Total-Variation-Diminishing methods for an improved treatment of steep
gradients. 

{\bf Thomas Wolf} presented his symbolic program CRACK for solving
partial differential equation. As examples, he demonstrated its use on
the Killing equations and on the PDE system which determines
infinitesimal symmetries of a 3rd order ODE that resulted in the
course of solving Einstein's field equations. 

Exotic smoothness and spacetime was the topic of {\bf Carl Brans}.  He
convincingly demonstrated that the smoothness property even of
topologycally trivial ${\bf R}^4$ are much more involved than many
relativists might have thought before. While he did not show colored
or animated visualization videos, he impressed the audience with the
information that all calculations for his lecture had been done on his
wristwatch, or rather on a cluster of $10^4$ such watches.

Wednesday afternoon, a {\bf hike} to a nearby village had been
scheduled. Joachim Debrus, the center's administrator, provided us
with a detailed coordinate system and warned us about various black
holes we might encounter along the way. Nevertheless, the experiment
showed that the combined expertise of so many theoretical relativists
did not suffice to find the shortest path from the center to our
destination. At least, nobody was captured by a black hole, so
everybody eventually made it to the restaurant overlooking the Rhine
river where coffee and cake was provided to revive our spirits after
this exhausting adventure. A boat had been chartered to bring us back
to Bad Honnef. It had about the size of a bathtub, but due to a
favorable spacetime distortion everybody found room on board. Despite
the sometimes rather daring route which our captain chose between the
larger tourist boats and the many freight ships, we arrived safely in
Bad Honnef.

{\bf Beverly Berger} explained how cosmological singularities are
investigated numerically, using symplectic integration to handle the
constraints. She demonstrated the use of Mathematica to generate
FORTRAN code for Einstein's equation form a spatially differenced
variational principle, and presented animated videos to visualize
her results.

{\bf Heinz Herold} examined rotating and oscillating neutron stars,
using a formulation equivalent to a minimum surface problem for the
stationary, axially symmetric equilibrium state. He discussed the
influence of the equation of state, central density, and angular
velocity of the star. He also presented a new slicing condition for
the ADM (3+1) formalism, based on the condition of constant mean
extrinsic curvature, and compared this condition to the familiar
maximal slicing and harmonic slicing conditions.

Scientific visualization in a relativistic context requires new
concepts and techniques. {\bf Hans-Peter Nollert} discussed
ray-tracing algorithms for special and general relativity, and how
they can be incorporated into `conventional' ray-tracing software. The
second lecture covered the visualization of two-dimensional surfaces
by embedding them in three-dimensional Euclidean space, either solving
a system of differential equations or constructing a wire frame
representation.

A massive, self-gravitating complex scalar field can form a stable a
boson star. {\bf Franz Schunck} showed how the corresponding rotating
configuration can be determined numerically, with regular energy
density and Tolman mass.

{\bf Pablo Laguna} discussed alternatives to the commonly used finite
difference methods in Numerical Relativity, such as spectral methods
and finite elements for solving the construction of initial data. In
the second part of his lecture, he addressed the question of matter
evolution in curved spacetimes, using particle-mesh techniques or
smoothed particle hydrodynamics in curved space.

{\bf Anton van de Ven} attacked the two-loop calculation for
perturbative quantum gravity using FORM, a relatively new computer
algebra program. He also gave a preliminary discussion of the
feasibility of a three-loop computation in supergravity. 

What happens at the center of our galaxy? {\bf R. Genzel} concluded
the school by presenting the fascinating detective's work that
observers do to figure out what's going on. While the luminosity of
the central few parsecs appears to be dominated by a cluster of hot
starts, there is now firly compelling evidence for a central dark
mass, consisting most likely of a million-solar-mass black hole.

Almost as important as the lectures were the informal discussions on
physics and trivia conducted in the evenings (and nights) in the
center's {\bf wine cellar}. A good selection of German beer as well as
German and French wines greatly facilitated these discussions. Some
participants, accidentally grabbed the non-alcoholic beer variety, and
were subsequently amazed at how much beer they could consume without
feeling any adverse effects.

The proceedings of the school will be published by Springer in early
1996. 

This year's school emphasized computer methods used in relativity. 
Next year, from August 19 -- 23, 1996, a similar school,
titled ``Relativistic Astrophysics'', will
concentrate on the results of relativistic studies, in particular with
respect to astrophysics.

\vfill\eject

\centerline{\bf Fifth Annual Midwest Relativity Conference}
\medskip
\centerline{Joseph D. Romano, University of Wisconsin--Milwaukee}
\centerline{romano@csd.uwm.edu}
\bigskip
\medskip

The Fifth Annual Midwest Relativity Conference was held on Friday,
November 10th and Saturday, November 11th at the University of 
Wisconsin--Milwaukee.  
Approximately seventy-five participants braved the cold, snowy 
weather to hear over fifty talks on a wide variety of topics, ranging 
from whether macroscopic traversable wormholes exist to the effect
of the accretion disk gap on the maximum angular velocity of relativistic
stars.

On Friday morning, Jorma Louko opened the conference with a talk on
complex actions in two-dimensional topology change.  
This was followed by a series of talks by Leonard Parker, Yoav Peleg, 
and Sukanta Bose on black hole evaporation, unitary evolution, and the 
Hamiltonian thermodynamics of 2D dilatonic black holes.
Eanna Flannagan and Steve Harris then described null spacetime 
singularities and the categoricity of casual boundary constructions, 
respectively.

Robert Mann, Kevin Chan, and Jolien Creighton reopened the discussion 
of black hole spacetimes by giving talks on the cosmological production 
of charged black hole pairs, the ADM mass for the GHS string black hole, 
and quasi-local thermodynamics of dilatonic black holes.
Steve Winters-Hilt followed by describing the Hamiltonian thermodynamics
of the Reissner-Nordstrom-anti-de-Sitter black hole, and John Friedman
talked about spherically symmetric mini-superspace reductions.
David Garfinkle and Comer Duncan closed the morning session with
back-to-back talks on numerical investigations of Choptuik scaling
in $n$ dimensions.

The Friday afternoon session began with talks by Dieter Brill and
Alan Steif on 2+1-dimensional multi-black hole geometries.
Matt Visser and Thomas Roman then described Lorentzian wormholes and 
asked the question whether macroscopic traversable wormholes exist.
Neil Cornish explained chaos and fractals in relativistic dynamics, 
and Tanmay Vachaspati talked about reproducing the standard model with
charges replaced by magnetic monopoles.
Gilad Lifschytz ended the first half of Friday afternoon session with
a talk on black hole thermodynamics from quantum gravity.

The last session on Friday afternoon was devoted to topics involving
numerical relativity.
Paul Casper described a numerical simulation of black hole formation
from collapsed cosmic string loops, while Mark Miller talked about the 
status of Regge calculus for numerical relativity calculations.
Spectral methods for numerical relativity and for numerical investigations 
of cosmological singularities were the topics of talks by Lawrence Kidder 
and Beverly Berger.
Nikolaos Stergioulas discussed the effect of the accretion disk gap
on the maximum angular velocity of relativistic stars, and Steve Brandt 
and Karen Camarda described initial data for rotating black holes, and 
3D numerical simulations of colliding black hole spacetimes.

Saturday morning began with talks on quantum field theory in curved
spacetime.
Robert Wald described recent work on quantum field theory in spacetimes 
with compactly generated Cauchy horizons, including a general theorem
that quantum fields have singular behavior at Cauchy horizons.
Adam Helfer and Rhett Herman discussed the stress-energy operator and 
the renormalization of the charged scalar field, respectively.
Theodore Quinn gave an axiomatic approach to radiation reaction in
curved spacetimes, and Alpan Raval talked about a stochastic theory of
accelerated quantum detectors.
Gerald Graef described semi-classical fluid spheres, and Larry Ford
told us about metric and light cone fluctuations in quantum gravity.

Asymptotic properties of spacetimes were also a topic of discussion,
with Shyan-Ming Perng giving a definition of a new conserved quantity at
spatial infinity, and Vivek Iyer describing a way of `deriving' a Bondi 
mass for diffeomorphism invariant theories.
Edward Glass followed with a talk about quasi-local mass, angular 
momentum, and the Penrose equation, and Elihu Lubkin described Lorentz 
transformations as seen in the planetarium sky.
Russel Cosgrove and Richard Epp ended the morning session 
with talks on consistent evolution for different time slicings
in quantum gravity, and the symplectic structure of GR in the (2,2)
formulation.

The Saturday afternoon session began with two talks involving the 
cosmological constant. 
Steve Leonard started by making a case for a positive cosmological
constant, while John Norbury described the quantum tunneling constraint
on a decaying cosmological constant.
Scott Koranda and Nicholas Phillips then gave talks on an absolute
minimum for the rotational period of gravitationally bound stars with
an equation of state constrained only by causality; 
and on texture cosmology and the cosmic microwave background.
Jorge Pullin described the use of linearized theory to understand the
head-on collision of two black holes, and Edward Schaefer proposed an
interpretation of GR based on an alternative definition of the constancy
of the speed of light.

The final session on Saturday afternoon was devoted primarily to
experimental relativity and gravity wave experiments.
Catalina Alvarez began with a talk on the equivalence principle and 
$g$-2 experiments.
Eric Poisson then told us ``the good news and bad news'' of post-Newtonian
approximations, while Liliana Simone and Alan Wiseman described the
convergence properties of post-Newtonian expansions for gravitational
waves, and the effect of equations of state on the inspiral of coalescing 
compact binaries.
James Geddes discussed high energy cosmic rays, and Daniel Holz 
concluded the conference with a talk on line-emitting accretion disks
surrounding Kerr black holes.

Finally, Comer Duncan volunteered to host the Sixth Annual Midwest 
Relativity Conference, to be held at Bowling Green State University
sometime in November, 1996.  
Stayed tuned for more information.

\vfil
\eject
\centerline{\bf Volga-7 '95}
\medskip
\centerline{Asja Aminova, Kazan State University and Dieter Brill, 
U of MD}
\centerline{aminova@phys.ksu.ras.ru, brill@umdhep.umd.edu}

\bigskip
\medskip

In the year of the 85th anniversary of the birthday of 
Prof.~A.~Z.~Petrov 
(1910-1972) a summer school dedicated to his memory was organized by 
the  
Chair  of  General  Relativity and Gravity, which he founded in 1961 
at 
Kazan State University. Held at Borovoye Matyushino (Kazan) from
June 22 to July 2, 1995, it was the 7th International Summer School 
on 
recent problems in theoretical and mathematical physics. Sponsors 
were 
the Ministry of Science of the Russian Federation, the Russian 
Foundation 
for Basic Research, and the Government of Republic of Tatarstan. 

The general purpose of this School is to create favourable conditions 
for 
involving young scientists and students in current scientific 
research, 
and to strengthen international scientific cooperation. 
The School's  71 participants came from 7 countries (Belgium, Czech 
Republic, 
Lebanon, The Netherlands, Russia, Ukraine, USA), and two
thirds were young scientists, postdocs, or graduate students. The 
languages 
of the School were Russian and English, with Russian predominating. 
This
presented an interesting challenge to the non-Russian-speaking 
participants
and to the organizers. Fortunately at this conference each
non-Russian-speaker
could have his personal simultaneous translator. Some lecturers 
solved the
language problem by speaking in Russian, but using overhead 
transparencies 
written in English. This was quite successful and is recommended for 
future
conferences with two official languages.

Three topics were presented as main lecture courses: \hfill\break
\line{Aspects of
Analyticity (D. Brill, Maryland, USA) \hfill} 
\line{Quantum Gravity (J. Buchbinder,
Tomsk, Russia)\hfill} 
\line{Elements of Mathematical Apparatus
of Quantum Field Theory\hfill}
\line{\hskip 5cm(Ju. Soloviyev, Moscow, Russia)\hfill} 

In addition we were treated to a variety of 90-minute seminars:
\hfill\break  
A. Aminova (Kazan, Russia) --- A. Z. Petrov as Scientist, Teacher and
Scientific Leader \hfill\break  
V. Bagrov (Tomsk,  Russia),  V.  Belov (Moscow,  Russia),  A. Trifonov
(Tomsk, Russia) ---  \break \indent
The Complex WKB Method \hfill\break  
\line{K. Bronnikov (Moscow, Russia) --- Insular 
Configurations in
Higher-Dimensional Gravity\hfill}
\line{K. Krasnov (Kiev, Ukraine) --- Nonperturbative Quantum Gravity 
and
Loop Expansion\hfill}
\line{M. Missarov (Kazan, Russia) --- Exactly Solvable Fermion 
Hierarchical
Models\hfill}
S. Stepanov (Vladimir, Russia) --- Irreducible Representations  of
the Orthogonal  Group \break \indent
and a Geometric Theory of Gravity \hfill\break  
\line{M. Welling (Utrecht, Holland) --- (2+1)-Dimensional 
Gravity\hfill}
\line{S. Chervon (Ul'yanovsk, Russia) --- Chiral Nonlinear $\sigma$-
Models
and Cosmological Inflation\hfill}
\line{I. Tsyganok (Vladimir, Russia) --- On 
Developable
Vector Fields\hfill}
\line{J. Steyaert (Louvain-la-Neuve, Belgium) --- On the 
Detection of Tachyons \hfill}  
\eject

These lectures and Seminars will be published in Russian and in
English as 
Proceedings of the Summer School and will be available from the
organizers (email enquiries to aminova@phys.ksu.ras.ru)

The name of the school recalls the magnificent river Volga that 
dominates
the geography of the Kazan region, and indeed the School was held at 
a camp 
within a forest reserve on the banks of the Volga. Participants were 
housed 
in rustic A-frames and huts nestled among pine trees. A large 
geodesic dome 
provided conference space and dining facilities. After intensive
discussions the fine weather invited recreation at the camp's volley- 
and basketball grounds, tennis courts, and beaches. There was
plenty of opportunity for interaction in discussion
as well as sports, and for typical Russian experiences such as the 
Sauna and
the Tatar national competition ``Sabantuy."  Organized events and 
excursions
included a visit of the City of Kazan and its Kremlin, a meeting with 
the
next generation of physicists and  mathematicians  at the Summer 
School ``Quant" for schoolchildren, a boat trip  to  the  island
Sviyazhsk (a notable 16th century architectural-religious  memorial), 
a talent show, a banquet with numerous toasts and dancing,
a concert by the University Chamber Orchestra conducted  by  Rustem
Abyazov with the  famous  violinist  Irina Bochkova (Moscow), and the
traditional farewell bonfire on the bank of the Volga.

\end

\end

\bye